\documentstyle[prl,aps,epsf]{revtex}

\begin{document}

\bibliographystyle{prsty}

\draft

\twocolumn

\title{Evidence for growth of collective excitations in glasses at low
temperatures} 

\author{Douglas Natelson, Danna Rosenberg, D.D.~Osheroff}

\address{Department of Physics, Stanford University, Stanford, CA94305-4060}

\date{\today}

\maketitle

\begin{abstract}
We present new data on the nonequilibrium acoustic response of 
glasses to an applied dc electric field below 1K.  When compared with
the analogous dielectric response of the same material, the acoustic
data show, within experimental precision, identical dependence on the
perturbing field, but stronger temperature dependence.  These
data are difficult to reconcile with simple generalizations of
the dipole gap model of two-level system (TLS) dielectric response,
unless we assume that as $T$ is decreased, interaction-based TLS
collective effects increase.
\end{abstract}

The equilibrium thermodynamic, acoustic, and dielectric properties of
amorphous materials at low temperatures are described well by the
tunneling two-level system (TLS) model \cite{standardTLS}.
Given a large number of independent TLS with a broad distribution of
energy asymmetries $\Delta$ and tunneling parameters $\Delta_0$
($P(\Delta, \Delta_0) = P_0/\Delta_0$, where $P_0$ is a contant), the
model predicts a specific heat $C \sim T$ and a thermal conductivity
$\kappa \sim T^2$, in fair agreement with the universal behavior noted
by Zeller and Pohl \cite{Pohl:71}.  TLS reach thermal equilibrium with
the phonon bath by coupling to local strain fields $u$, which shift
asymmetries by $\gamma \cdot u$, where $\gamma \approx$ 1 eV.  Via
this coupling, the TLS contribute to the equilibrium sound speed and
internal friction \cite{Jackle:72}.  If the TLS have intrinsic
electric dipole moments of magnitude $p$, the TLS contribution to the
equilibrium dielectric response \cite{Anthony:Anderson:79} is strictly
analogous.

The equilibrium sound speed $v$ has a maximum at a temperature $T_{\rm
m} \approx 50$ mK.  For $T > T_{\rm m}$, the ``relaxational regime'',
TLS behavior is dominated by single-phonon relaxational processes,
while for $T < T_{\rm m}$, the ``resonant regime'', resonant quantum
tunneling of the TLS dominates their response to an ac strain field.
The real part of the dielectric response $\epsilon'$ behaves
analogously, but with an overall sign difference.  In both $v$ and
$\epsilon'$ the crossover temperature $T_{\rm m}$ at measuring
frequency $\omega$ is set by the TLS-phonon coupling, $T_{\rm m} =
\left(\frac{\omega}{A}\right)^{\frac{1}{3}}$, where $A \propto
\gamma^{2}$.

Recent measurements \cite{Salvino:94,Rogge:96,Rogge:97} of the
nonequilibrium dielectric response of glasses are consistent with a
model \cite{Burin:95} in which weak dipolar TLS-TLS interactions
produce a ``dipole gap'' in the TLS density of states near zero local
electric and strain fields.  The gap is necessary to ensure the stability
of the ground state, and has been discussed elsewhere
\cite{Kirkpatrick:77} in connection with spin glasses.

We concentrate on ``sweep'' experiments,
in which an applied electric field, $E_{\rm dc}$, is slowly varied in
time as a triangle wave, and the nonequilibrium dielectric response is
measured as a function of that field.  This response shows a minimum
at zero applied bias (a ``zero bias hole''), which becomes sharper and
deeper as the temperature is lowered.  These sweep data are
interpreted as an observation of the proposed gap in the TLS density
of states.  While the external field is swept, TLS with short
relaxation times compared to the period of the sweep partly fill the
zero bias hole, an effect which is most important when $T \lesssim 15$
mK.  Slower TLS continue to exhibit a depressed density of states at
the field in which the sample was cooled.  When $T \lesssim 15$ mK,
hysteretic effects \cite{Rogge:97} at the extrema of the sweeps become
important, and for analysis these data are clipped to avoid edge
effects.

In a standard capacitance bridge measurement, a small ($\sim$1000 V/m)
ac electric field is used to probe the dielectric response.  For a
typical TLS electric dipole moment, $p \approx$ 1 Debye ($3.33 \times
10^{-30}$ C m), this corresponds to an oscillating energy shift
equivalent to $\approx 200 \mu$K.  Not all TLS have large electric
dipoles \cite{Schickfus:76,Laermans:77}.  For example, it has been
determined \cite{Golding:79} that OH$^{-}$ impurities in SiO$_2$ have
dipole moments of around 3.7 D and dominate $\epsilon'$, while the
intrinsic SiO$_2$ TLS have moments of 0.6 D.  The TLS dielectric
response is proportional to $p^{2}$, so dielectric measurements are
most sensitive to nearly degenerate TLS with substantial electric
dipole moments.  In contrast, the acoustic dispersion is influenced by
all TLS with elastic couplings, not just those with large $p$.

The motivation of the present work was to examine the acoustic
response to a perturbation which directly affects only the
dielectrically active TLS.  A related experiment, in which all TLS
were perturbed with an applied strain field while only the
dielectrically active subset was monitored, has been reported
elsewhere \cite{Rogge:97}.

To measure the acoustic response, torsional oscillators were
constructed from sample materials, consisting of a torsion bar 150
$\mu$m thick, 3 mm wide, and 3.5 cm in length.  A paddle of the same
material $150\mu$m thick, 1 cm in length, and 2.2 cm in width was
attached to the center of the torsion bar using a minute quantity of
epoxy.  The ends of each bar were clamped between copper blocks
mounted on an additional piece of the sample material to avoid strain
due to thermal contraction.  The bending length of the torsional
member was $\approx$ 1.5 cm. A gold ground plane was evaporated on
one side of the oscillator, and gold pads were deposited on the other
side of the torsion bar to provide the perturbing field, $E_{\rm
dc}$.  A NdFeB magnet roughly 0.5 mm$^{3}$ in volume was epoxied to
the center of the paddle, with its magnetization oriented in the plane
of the paddle but perpendicular to the axis of the bar.  The
oscillator, mounted on a dilution refrigerator, was driven
magnetically with a small superconducting coil and
sensed capacitively.

Two oscillators were constructed, one from a commercial microscope
coverglass\cite{coverglass}, and the other from a previously studied
\cite{Anthony:Anderson:79} lot of BK7, a standard optical glass.  The
natural frequency, $f_{0}$, of the coverglass oscillator at 4.2 K was
approximately 490 Hz, and the internal friction, $Q^{-1}$, as measured
by a plot of amplitude v. driving frequency, was $5 \times 10^{-4}$,
including clamping losses, not unusual for such materials.  For the
BK7 oscillator at 4.2 K, $f_{0} \approx$ 420 Hz, and the total
$Q^{-1}$ was 6.7 $\times 10^{-4}$.

Table I summarizes the equilibrium acoustic and dielectric responses
measured with these techniques.  As in previous equilibrium studies of
such materials \cite{Esquinazi:92,Classen:94,Rogge:96}, the internal
friction and dielectric losses were best described by an approximately
linear variation with $T$ for $T < T_{\rm m}$, in disagreement with
the standard tunneling model's prediction of a $T^{3}$ dependence.
Similarly, the measured ratios of slopes of $\frac{\delta v}{v}$ and
$\frac{\delta \epsilon'}{\epsilon}$ v. $\ln T$ for low and high
temperatures were closer to -1:1 than the predicted -2:1 for both
materials.

\begin{figure}
\epsfclipon
\epsfxsize=8.5cm
\epsfbox{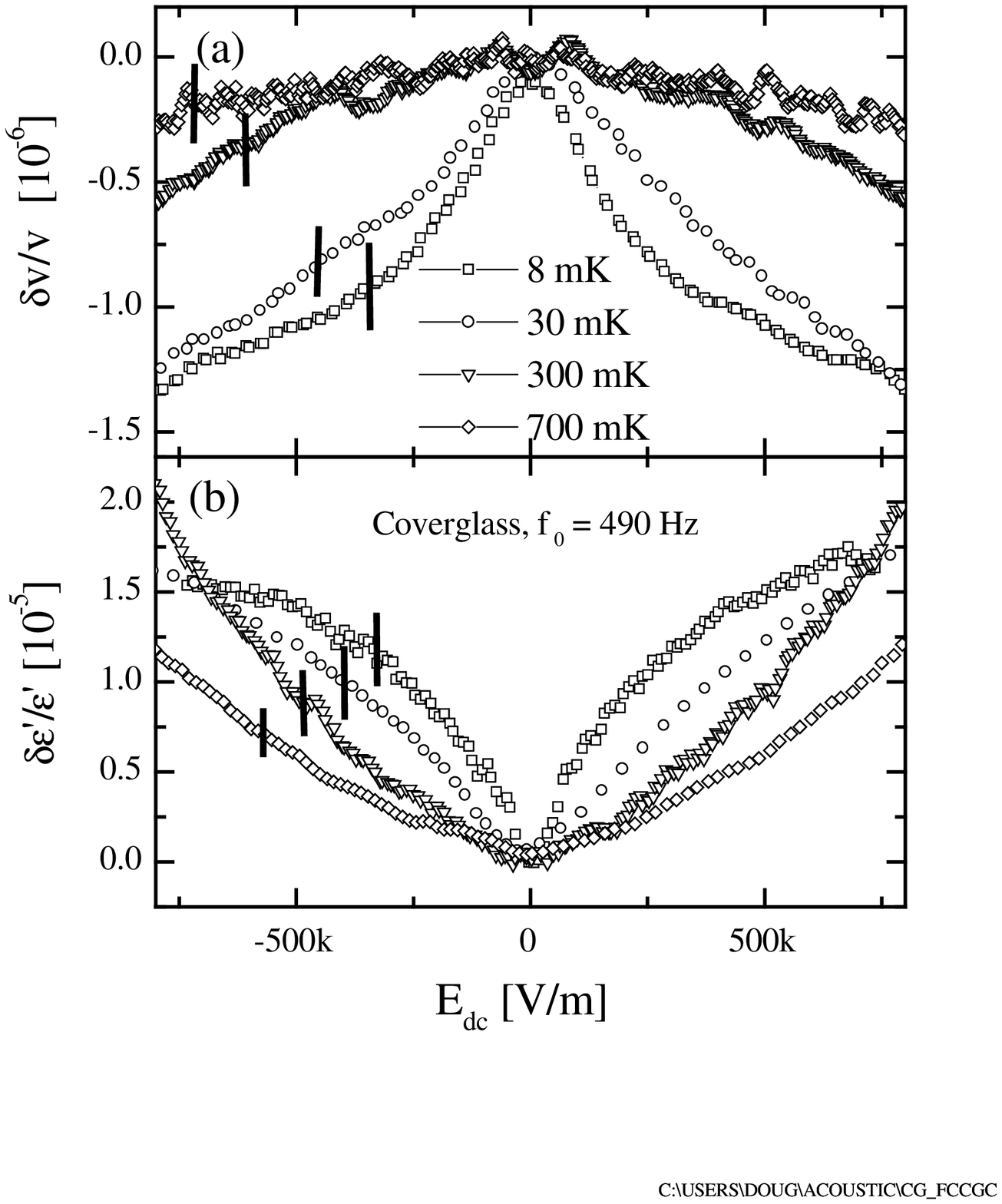}
\caption{Acoustic and dielectric response of coverglass $vs.$ applied
$E_{\rm dc}$ at various temperatures, as the field was swept at
1.3 kV/m$\cdot$s.  Measuring frequency was 490 Hz, and no excitation
or sweep rate dependence was observed.  The data have been
smoothed by adjacent averaging; statistical errors are indicated by
vertical bars.}
\label{swps.fig}
\end{figure}

Figure \ref{swps.fig}a presents the results of acoustic sweep
experiments on the coverglass sample.  Recalling that increased TLS
response corresponds to a lower sound speed, it is clear that a hole,
centered at zero applied dc electric field, whose shape and depth are
strongly temperature dependent, develops in the response.  These data
are particularly interesting when compared with analogous dielectric
measurements of the same material, shown in Fig. \ref{swps.fig}b.  The
qualitative behavior is the same, but the temperature dependence of
$\frac{\delta \epsilon'}{\epsilon'}$ is considerably smaller than that
of $\frac{\delta v}{v}$.  There is still a substantial gap in
the dielectric response at 700 mK, while the acoustic data 
show a much smaller response at that temperature.  The data for 
BK7 are qualitatively the same.  There was no significant drive level
dependence to either the acoustic or dielectric nonequilibrium response
(at acoustic strain amplitudes up to $1 \times 10^{-6}$ and ac
measuring field amplitudes up to 2 kV/m).  All the curves in
Fig. \ref{swps.fig} are taken well after the initial
transient response.  We assume that the filling of the zero bias hole
during the sweep, which is only significant below 15 mK,
is identical for acoustic and dielectric sweeps, an assumption
justified by the observation that TLS relaxation rates are 
the same for both strain and electric field perturbations \cite{Rogge:97}.

The dipolar gap theory developed to explain the dielectric effects
\cite{Burin:95} assumes that all TLS have a dipole moment of magnitude
$p$ and that the response at short perturbation times is due to a
field-dependent TLS density of states, $\delta P(\Delta,
\Delta_{0},E_{\rm dc},T) \equiv P(\Delta, \Delta_{0}, E_{\rm dc},T) -
P(\Delta, \Delta_{0},E_{\rm dc} = 0,T)$.  Define a function $g$ such
that the equilibrium $\delta \epsilon'(T,\omega)/\epsilon$ for a
single TLS is proportional to $g(\Delta, \Delta_{0},\omega,T)$.  The
dipole gap response is then:
\begin{eqnarray}
\left(\frac{\delta \epsilon'}{\epsilon}\right)(E_{\rm
dc},\omega,T) & = & \frac{2 p^2}{3 \epsilon_{0} \epsilon} \int \int \delta
P(\Delta, \Delta_{0}, E_{\rm dc},T) \cdot \nonumber \\ 
& & g(\Delta, \Delta_{0},\omega,T) d\Delta~d\Delta_{0} \nonumber \\
& = & \frac{2 p^2}{3
\epsilon_{0} \epsilon} h(\frac{pE_{\rm dc}}{k_{\rm B}T}, \frac{A
T^{3}}{\omega}).
\label{eq:capgap}
\end{eqnarray}
The function $h$ defined above contains the $E_{\rm dc}$ and $T$
dependences of $\delta P$ and the $T$ and $\omega$ dependence of the
equilibrium dielectric response, and depends logarithmically on
$\frac{A T^{3}}{\omega}$ \cite{Burin:95}.

\begin{figure}
\epsfclipon
\epsfxsize=8.5cm
\epsfbox{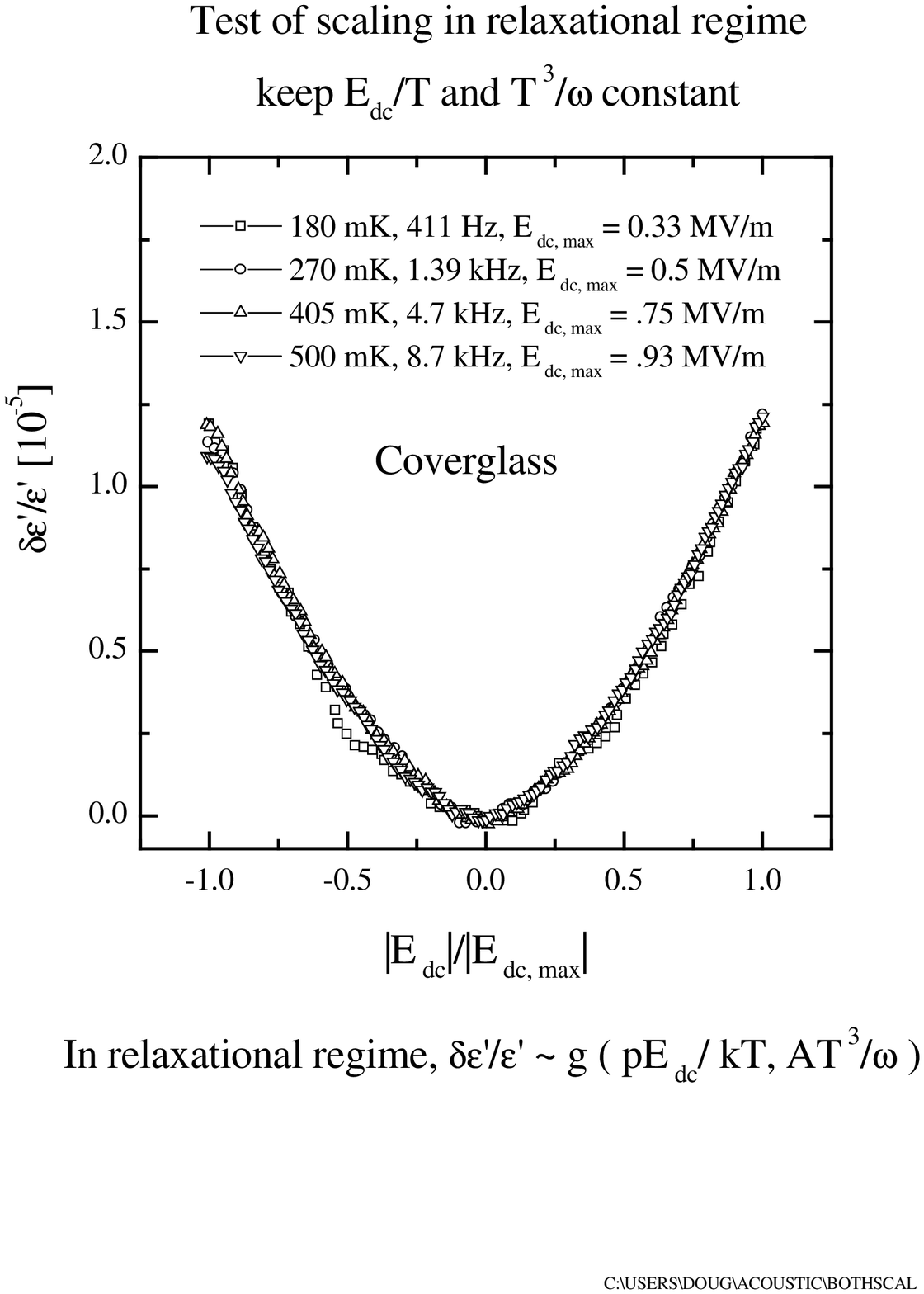}
\caption{Test of Eq. (3) on the coverglass sample.  Dielectric sweeps
at various $E_{\rm dc, max}, \omega,$ and $T$, with $E_{\rm dc,
max}/T$ and $T^{3}/\omega$ constant, showing that only these
parameters determine the nonequilibrium dielectric response.}
\label{scale.fig}
\end{figure}

To test the applicability of (\ref{eq:capgap}), we performed a series
of sweeps in the accessible domain of $E_{\rm dc}, \omega,$ and $T$,
such that $\frac{E_{\rm dc, max}}{T}$ and $\frac{T^{3}}{\omega}$ were
kept fixed.  The results for the coverglass sample are shown in Figure
\ref{scale.fig}.  Again, the BK7 exhibited quite similar behavior.
Within the precision of the data, (\ref{eq:capgap}) accurately
describes the parametric dependence of the nonequilibrium dielectric
response.  Unfortunately, we cannot easily vary $f_{0}$ of the
torsional oscillators to check for this scaling in the acoustic
response.

We can generalize the above to the case where only a fraction $f$ of
the TLS have electric dipoles $p$.  One would then expect the acoustic
response to $E_{\rm dc}$ to be:
\begin{equation}
\left(\frac{\delta v}{v}\right)_{\rm noneq}(E_{\rm dc}, \omega, T)=
\frac{f \gamma^2}{\rho v^2} h(\frac{pE_{\rm dc}}{k_{\rm B}T}, \frac{A
T^{3}}{\omega}),
\label{eq:simplemodel}
\end{equation}
where we have assumed that the dielectrically active TLS have the same
$\gamma$ as the rest of the TLS.  Eqs. (\ref{eq:capgap}) and
(\ref{eq:simplemodel}) differ only by a constant factor, and thus this
simple treatment cannot explain the differing $T$ dependences of the
acoustic and dielectric sweeps seen in Fig. \ref{swps.fig}.

\begin{figure}
\epsfclipon
\epsfxsize=8.5cm
\epsfbox{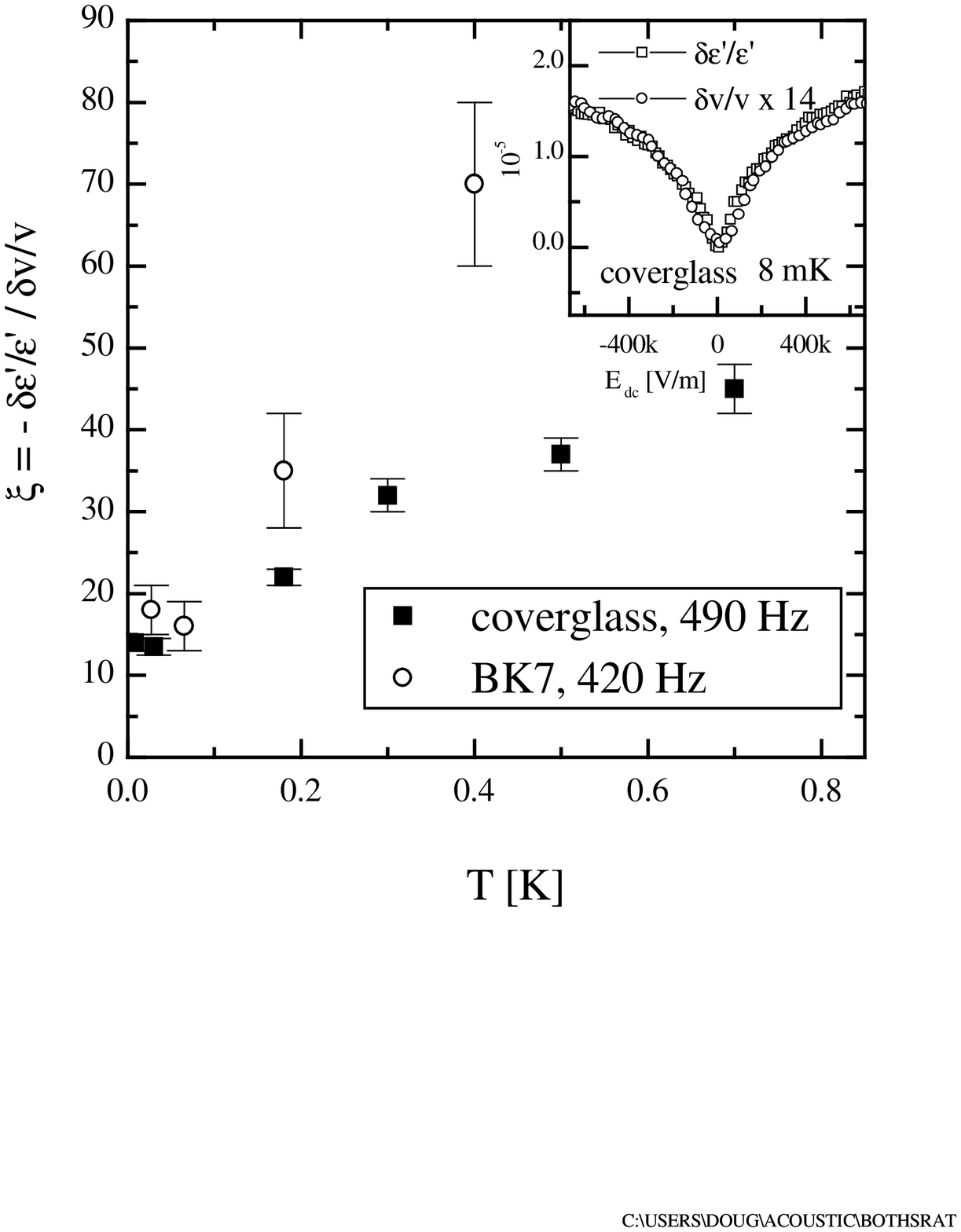}
\caption{The scale factor $\xi$ at various temperatures for the two
samples.  A curve of $\frac{\delta v}{v}$ vs. $E_{\rm dc}$ at
temperature $T$, when multiplied by $\xi(T)$, falls on top of the
dielectric data at that same temperature; see inset for an example.
Error bars estimate the deviation in $\xi$ before one obtains visible
systematic deviations in the scaling of the smoothed data.}
\label{srat.fig}
\end{figure}

Within the statistical precision of the data, the dc field dependences
of the nonequilibrium acoustic and dielectric responses share the same
functional form; for example, see the inset in Fig. \ref{srat.fig}.
For each $T$ a single multiplicative factor $\xi$ can make the
acoustic data fall on top of the dielectric data, and the residuals
show no systematic trends as a function of bias field.  The
temperature dependence of $\xi$ is shown in Fig. \ref{srat.fig} for
both samples, and is substantially larger in BK7 than in coverglass.
Identical field dependences suggest that TLS with electric dipoles of
a single magnitude are responsible for both effects, but the simple
model of Eq. (\ref{eq:simplemodel}) isn't correct, since it predicts
$\xi$ to be constant.  We also see that at the lowest temperatures the
factor $\xi$ saturates to a constant value, although both the
equilibrium and nonequilibrium acoustic and dielectric responses
continue to change.

We now consider attempts to explain the behavior in
Fig. \ref{swps.fig} by modifying the TLS distribution used in
Eqs. (\ref{eq:capgap}) and (\ref{eq:simplemodel}).  In particular,
distributions of the TLS coupling constants $p$ and $\gamma$ might
complicate matters \cite{Burin:pricom:97}.  Within the dipole gap
model, one possible source of the relative temperature dependence seen
in Fig. \ref{swps.fig} could be differing relaxation parameters
$A$ for the TLS responsible for the acoustic and dielectric responses.
However, this would be inconsistent with Table I, which shows
that $T_{{\rm m},v} \approx T_{{\rm m},\epsilon}$.

Given this constraint on $A$, the differing $T$ dependences of
acoustic and dielectric responses in this model would have to come
from different dependences on $E_{\rm dc}/T$, implying that the
$E_{\rm dc}$ dependences should be different as well.  To be
consistent with the data, any distribution of $p$ proposed
to explain the temperature dependence of $\xi$ would have to lead to a
difference in $E_{\rm dc}$ dependences too small to be resolved in
this experiment.

We can also extend the model by explicitly accounting for the
$T$ dependence of the interactions between TLS.  The effective
range of TLS interactions depends on $T$, since there is
some length scale beyond which the interaction energy is small
compared to $k_{\rm B} T$.  Consider a particular TLS.  We may define
a volume centered on that TLS within which the interaction energy is
greater than $k_{\rm B} T$.  For dipolar interactions, that volume
grows as $T^{-1}$ with decreasing temperature.

We propose that the fraction of TLS which are directly perturbed
substantially by our bias field is relatively small.  At high
temperatures, interactions are short-range, and our acoustic data show
a very small response because such a small fraction of acoustic TLS
are perturbed.  With decreasing $T$, however, the growth of the
interaction range allows the directly perturbed TLS to couple to an
increasing number of dielectrically inactive TLS.  The identical field
dependences would be expected, since the number of dielectrically
inactive TLS affected via interactions would be proportional to the
number of TLS which are directly perturbed by $E_{\rm dc}$.  At
sufficiently low temperatures, the typical range of interactions is
comparable to the average spacing between dielectrically active TLS,
and further decreases of $T$ have little effect on the number of
indirectly perturbed TLS.  Thus one would expect the temperature
dependence of $\xi$ to saturate at low $T$, as is seen.  Note that the
dipole gap itself continues to grow with decreasing temperature, as
thermal smearing is reduced.

While a quantitative model for such a mechanism does not yet exist, an
explanation of the difference between the acoustic and dielectric data
in terms of the growing effect of TLS-TLS interactions seems necessary
to explain the observed $E_{\rm dc}$ and $T$ dependences.  Indirect
support for interactions as a modification of the standard tunneling
model has been seen in the temperature dependence of the dielectric
loss tangent \cite{Rogge:96} and the internal friction
\cite{Esquinazi:92,Burin:Kagan:95}, and in recent spectral hole
burning experiments \cite{Maier:96,Neu:97}.  The current work and
related experiments \cite{unpubl:98} provide important evidence that
interaction-based collective effects grow as the temperature is
decreased below 1 K.  As various models
\cite{variousmodels} are
proposed which incorporate interacting TLS to explain the
universality of the low temperature properties of these materials,
experiments which directly examine interaction effects are of
increasing importance.

The authors wish to acknowledge valuable discussions of their data and
the theoretical background with Alexander Burin, and to thank
Christian Enss for his insights.  D.N.  acknowledges fellowship
support from the Fannie and John Hertz Foundation. This work was
supported under Department of Energy grant DE-FG03-90ER45435.


\begin{table}
\onecolumn
\caption{Equilibrium dielectric and acoustic properties of the
capacitors and oscillators.  Dimensionless coefficients $C_{\rm d}
\equiv \frac{2 P_{0}p^{2}}{3 \epsilon_{0}\epsilon}$ and $C_{\rm a}
\equiv \frac{P_{0}\gamma^2}{\rho v^2}$ are obtained from the $T >
T_{\rm m}$ data up to 700 mK in the linear drive regime ($u$ in $v$
measurements $\le 10^{-7}$, $E_{\rm ac}$ in $\epsilon'$ measurements
$\le$ 200 V/m), assuming the standard TLS model.}
\begin{tabular}{c c c c c c}
Sample & freq. & $C_{\rm d}$ & $T_{\rm m,d}$ & $C_{\rm a}$ & $T_{\rm
m,a}$\\
\hline
BK7 & 420 Hz & $1.51 \pm .15 \times 10^{-3}$ & $49 \pm 2$ mK & $3.20
\pm .3 \times 10^{-4}$ & $50 \pm 2$ mK \\
coverglass & 490 Hz & $1.69 \pm .15 \times 10^{-3}$ & $51 \pm 2$ mK & $3.60
\pm .3 \times 10^{-4}$ & $51 \pm 2$ mK \\
\end{tabular}
\end{table}

\end{document}